\documentclass
[%
reprint,
 showpacs,preprintnumbers,
 amsmath,amssymb,
 aps,
 pre,
 floatfix,
 longbibliography
]{revtex4-1}

\usepackage{booktabs}
\usepackage{array}
\newcolumntype{L}[1]{>{\raggedleft\arraybackslash}p{#1}}
\AtBeginDocument{
\heavyrulewidth=.08em
\lightrulewidth=.05em
\cmidrulewidth=.03em
\belowrulesep=.65ex
\belowbottomsep=0pt
\aboverulesep=.4ex
\abovetopsep=0pt
\cmidrulesep=\doublerulesep
\cmidrulekern=.5em
\defaultaddspace=.5em
}

\usepackage{color}
\usepackage{xcolor}
\usepackage{graphicx}% Include figure files
\usepackage{dcolumn}% Align table columns on decimal point
\usepackage{bm}% bold math
\usepackage{cancel}
\usepackage[utf8]{inputenc}
\newcommand{\dpar}[2]{\frac{\partial #1}{\partial #2}}

\begin{document}

%\preprint{APS/123-QED}

\title{Diffusion in binary mixtures: an analysis of the dependence on the thermodynamic factor}

\author{M. Di Pietro Mart\'inez}
%\email{mdipietro@ifimar-conicet.gob.ar}
\author{M. Hoyuelos}
\email{hoyuelos@mdp.edu.ar}

\affiliation{Instituto de Investigaciones F\'isicas de Mar del Plata (IFIMAR -- CONICET)}
\affiliation{Departamento de F\'isica, Facultad de Ciencias Exactas y Naturales,
             Universidad Nacional de Mar del Plata.\\
             De\'an Funes 3350, 7600 Mar del Plata, Argentina}

\date{\today}% It is always \today, today,
             % but any date may be explicitly specified

\begin{abstract}
We study the diffusion process in binary mixtures using transition probabilities that depend on a mean-field potential. This approach reproduces the Darken equation, a relationship between the intrinsic and the tracer diffusion coefficients, $D_A$ and $D_A^*$, through the thermodynamic factor $\Phi$ (a function of the derivative of the activity coefficient against molar fraction). The mean-field approach allows us to go beyond the Darken equation and separately specify the dependence of $D_A$ and $D_A^*$ on the thermodynamic factor. We obtain that $\Phi$ appears in the expression for $D_A^*$, but the intrinsic diffusivity $D_A$ turns out to be independent of $\Phi$. Experimental results taken from the literature on diffusion in metal alloys are consistent with this theoretical prediction.
\end{abstract}


\maketitle

%##################################################################################
\section{Introduction}

Diffusion in solids is of crucial importance in material science. It is closely related to the durability of a compound, the conduction and transport properties, and it is applied to, for example, the design of new materials, the preparation, processing and subsequent heat treatment for hardening and toughening, etc.
The physics of the diffusion process is complex.
It involves many-body dynamics, intricate interactions between different types of atoms and holes, and there are a number of different aspects that impacts in the diffusion process,
such as the presence of vacancies, impurities, if the alloy is diluted or concentrated,
the temperature, the pressure,
the melting properties, activation energies, elastic constants, etc.
The detailed effect of each of these ingredients, and more, can be found in, e.g.,  Refs.~\cite{mehrer,paul,paul2,shewmon,sohn}.

The mean-field approach permits to analyze a complex dynamic and reduce it to a problem where the analytical calculations can be attainable and usually complete.
The aim of this work is to find an analytical expression for the diffusion coefficients in binary mixtures in the framework of mean-field theory, isolating the key ingredients to reproduce the observed dynamics in metal alloys. The main purpose is to separately specify the dependence of the intrinsic and tracer diffusion coefficients on the thermodynamic factor.

This article is organized as follows. Firstly, in Sec.~\ref{sec:theory}, we condense the background theory of diffusion introduced initially by Darken. In Sec.~\ref{sec:mean}, we present the details of the mean-field approach and how it is related to the experimental system of a binary mixture.
Subsequently, in Sec.~\ref{sec:theta}, we calculate analytically the relationship between mean-field parameters to ultimately find their dependence on the physical observables such as the thermodynamic factor and the activation energy.
In Sec.~\ref{sec:thermofactor} we conclude the theoretical analysis with the derivation of an expression for the intrinsic diffusivity as a function of the concentration. Finally, in Sec.~\ref{sec:exptest}, we test this expression fitting experimental data for diffusion in metallic alloys.
We close in Sec.~\ref{sec:conclusions} with the conclusions and discussion.

\section{Basic theory}
\label{sec:theory}

We briefly summarize the theoretical description of substitutional diffusion in solid binary alloys, originally proposed by Darken \cite{darken}; see also \cite[ch. 10]{mehrer}. Let us call $c_A$ and $c_B$ the molar concentrations of species $A$ and $B$; the total concentration is $c_T = c_A + c_B$. For simplicity, we consider spatial variations only along the $x$ axis, and we write the equations for species $A$; the corresponding equations for species $B$ are immediately obtained exchanging $A \leftrightarrow B$. The diffusion current respect to the crystalline lattice is
\begin{equation}
J_A = -D_A \dpar{c_A}{x}
\end{equation}
where $D_A$ is the intrinsic diffusivity for species $A$. There could be a net volume flux through a crystalline plane perpendicular to the $x$ axis. In the laboratory reference frame (in which the volume current is zero), such plane moves with the Kirkendall velocity, given by
\begin{equation}
v_K = -\nu_A J_A - \nu_B J_B,
\end{equation}
where $\nu_A$ and $\nu_B$ are the partial molar volumes; the total molar volume is $\nu_m = 1/c_T = N_A \nu_A + N_B \nu_B$, where $N_A=c_A/c_T$ and $N_B=c_B/c_T$ are the mole fractions; we also have that $ c_A \nu_A + c_B \nu_B=1$ and $\nu_A \, dc_A + \nu_B \,dc_B =0$. It can be seen that the current of species $A$ or $B$ in the laboratory reference frame, $J_{A,\text{lab}}$ and $J_{B,\text{lab}}$, have the same diffusion coefficient:
\begin{equation}
J_{A,\text{lab}} = J_A + c_A v_K = -\tilde{D} \dpar{c_A}{x}
\end{equation}
where
\begin{equation}
\tilde{D} = \nu_B c_B D_A + \nu_A c_A D_B
\label{interdif}
\end{equation}
is the interdiffusion coefficient.

The thermodynamic force that drives the particle current is the gradient of the chemical potential $\mu_A$, and the linear relationship that connects current and force is $J_A = B_A c_A \dpar{\mu_A}{x}$, where $B_A$ is the mobility of species $A$. Using the Einstein relation, $D_A^*=B_A R T$, where $D_A^*$ is the tracer diffusion coefficient and $R$ is the ideal gas constant, we have
\begin{equation}
J_A = \frac{D_A^* c_A}{R T} \dpar{\mu_A}{x}.
\end{equation}
Given the expression of the chemical potential in terms of activity coefficient $\gamma_A$,
\begin{equation}
\mu_A = \mu_A^0 + RT \ln (N_A\gamma_A),
\label{chempot}
\end{equation}
the following relationship between intrinsic and tracer diffusivity is obtained \cite[Sec.\ 10.3]{mehrer}:
\begin{equation}
D_A = D_A^* \frac{\nu_m}{\nu_{B}} \Phi,
\label{dark1}
\end{equation}
where
\begin{equation}
\Phi = 1 + \dpar{\ln \gamma_A}{\ln N_A}
\label{eq:Phi}
\end{equation}
is the thermodynamic factor. Using the Gibbs-Duhem relation, it can be shown that the thermodynamic factor is the same for species $A$ or $B$. Replacing \eqref{dark1} in \eqref{interdif}, we get the following simple equation for the interdiffusion coefficient
\begin{equation}
\tilde{D} = (N_A D_B^* + N_B D_A^*)\Phi.
\label{dark2}
\end{equation}
Equations \eqref{dark1} and \eqref{dark2} are called Darken equations. In his original derivation, Darken assumed constant total concentration, that means that $\nu_m=\nu_A=\nu_B$; in this particular case, Eq.\ \eqref{dark1} becomes $D_A = D_A^* \Phi$.

Substitutional diffusion is not possible without the presence of vacancies. For small enough vacancy concentration, the probability of finding a vacancy can be taken equal to its mole fraction $N_V$, and the (tracer or intrinsic) diffusivities should be proportional to $N_V$ (see \cite[Sec. 5.3]{paul})

%The tracer diffusivity depends on $N_V$ and on the migration energy per particle, $\Delta g_A$ or $\Delta g_B$, in the following way (see, e.g., \cite{paul})
%\begin{equation}
%D_A^* = d_A N_V e^{-\beta \Delta g_A}  \qquad A=A,B
%\end{equation}
%where $d_A$ is a concentration and temperature-independent prefactor that encompasses the Debye frequency, the lattice spacing and the correlation factor (see \cite[Sec. 5.3]{paul}). In the limit of small concentration, the tracer and intrinsic (or collective) diffusivities are equal to the free diffusion coefficient $D_{A 0}$; the vacancy mole fraction is $N_V^{A 0}$ and the migration energy is $\Delta g_{A 0}$, then
%\begin{equation}
%D_{A 0} = d_A N_V^{A 0} e^{-\beta \Delta g_{A 0}}.
%\end{equation}
%We can write the tracer diffusivity in terms of the free diffusion coefficient as
%\begin{equation}
%D_A^* = D_{A 0} \frac{N_V}{N_V^{A 0}} e^{-\beta(\Delta g_A - \Delta g_{A 0})}.
%\end{equation}

\section{Mean field approach}
\label{sec:mean}

We divide the system in cells of length $a$. The cell size should be much smaller than the characteristic length of the concentration inhomogeneities, so that the cell can be considered point-like, and, at the same time, much larger than the lattice spacing. We assume smooth enough spatial and temporal variations, so that the local thermal equilibrium approximation holds. We can write the transition probability from the cell with label $i$ to the neighboring cell $i+1$, as
\begin{equation}
W_{i,i+1}^A = P_A \exp \left[ -\frac{\beta}{2} \left( \theta_{A,i+1}  V_{A,i+1} + \theta_{A,i} V_{A,i} + \Delta V_A \right) \right]
\label{eq:transition}
\end{equation}
where $P_A$ is the jump rate of particles $A$; $V_{A,i}$ is the mean field potential for one particle in cell $i$; $\Delta V_A=V_{A,i+1} - V_{A,i}$; and $\theta_{A,i}$ is an interpolation parameter that determines if $W_{i,i+1}^A$ depends on the potential in the origin cell $i$, on the one in the target cell $i+1$, or on a combination of both. The mean field potential is a function of the number of particles $n_{A,i}$, and $V_{A,i}$ is an abbreviation of $V_A(n_{A,i})$.  Eq.\ \eqref{eq:transition} was proposed in \cite{suarez} as a general form of the Arrhenius formula that satisfies detailed balance. The starting point of the present calculations is the expression for the transition probability \eqref{eq:transition}. In the next paragraphs, we reproduce some results of Ref.\ \cite{dipietro2} for completeness.

The particle current of species $A$ for a given configuration is given by $n_i^A W_{i,i+1}^A - n_{i+1}^A W_{i+1,i}^A$. Taking the average on configurations and the continuous limit we obtain (see, e.g., Appendix A in Ref.\ \cite{dipietro2} for details)
\begin{equation}
J_A = - \Delta_A \, e^{-\beta \theta_A V_A}\left(\beta c_A \dpar{V_A}{x} + \dpar{c_A}{x} \right)
\label{corrJ}
\end{equation}
where $\Delta_A = P_A a^2$ has units of diffusion coefficient. Let us note that $\Delta_A$ may depend on position or concentration.

Comparing the zero current equilibrium solution for the concentration with the expression that comes from the chemical potential \eqref{chempot}:
\begin{equation}
c_A = c_T e^{(\mu_A - \mu_A^0 - RT \ln \gamma_A)/RT},
\label{eq:conc}
\end{equation}
we can obtain a relationship between the mean field potential and the activity coefficient \cite{dipietro2}
\begin{equation}
\beta V_A = \ln \frac{\gamma_A\, c_{B0}}{\gamma_{A 0}\, c_T},
\label{va1}
\end{equation}
where $c_{B 0}$ is the concentration of pure $B$, and $\gamma_{A 0}$ is the activity coefficient for $c_A\rightarrow 0$. In the limit of small concentration, the condition $V_A \rightarrow 0$ is satisfied. Replacing \eqref{va1} in \eqref{corrJ}, after some algebra (see Appendix A) we get
\begin{equation}
J_A = -\Delta_A  \, e^{-\beta \theta_A V_A}\, \Phi \, \frac{\nu_m}{\nu_{B}} \, \dpar{c_A}{x},
\label{eq:Jalpha}
\end{equation}
where we can identify the intrinsic diffusivity
\begin{equation}
D_A = \Delta_A  \, e^{-\beta \theta_A V_A}\, \Phi \, \frac{\nu_m}{\nu_{B}}.
\label{eq:intr}
\end{equation}

The transition probabilities can also be used to obtain the tracer diffusion coefficient $D_A^*$ through the evaluation of the mean square displacement in a short time interval $\Delta t$: $\langle (\Delta x)^2 \rangle = 2 D_A^* \,\Delta t$ (see Appendix B in Ref.\ \cite{dipietro2}). We obtain
\begin{equation}
D_A^* = \Delta_A  \, e^{-\beta \theta_A V_A}.
\label{eq:trac}
\end{equation}
Combining \eqref{eq:intr} and \eqref{eq:trac}, we recover the Darken equation \eqref{dark1}: $D_A = D_A^* \frac{\nu_m}{\nu_{B}} \Phi$.

As mentioned in the previous section, both diffusivities are proportional to the vacancy mole fraction $N_V$ for a substitutional alloy. This dependence is included in $\Delta_A$, since the jump rate $P_A$ between cells should also be proportional to $N_V$. The vacancy mole fraction is given by $N_V=e^{-\beta g_V}$, where $g_V$ is the vacancy formation energy. Besides vacancies, we should also include the effect of the migration energy $g_m^A$ for a particle of type $A$. The diffusivities are proportional to $e^{-\beta g_A}$, where $g_A = g_V + g_m^A$ is the activation energy (see \cite[Sec. 5.3.5]{paul}); for a binary alloy we have that $g_A$ (and $g_V$ and $g_m^A$) is a function of the molar fraction. We can write
\begin{equation}
\Delta_A = D_{A 0}\, e^{-\beta (g_A-g_{A 0})},
\label{eq:Delta}
\end{equation}
where $g_{A 0}$ and $D_{A 0}$ are, respectively, the activation energy and the diffusivity when $c_A\rightarrow 0$. In the limit of small concentration we have that $D_A = D_A^* = D_{A 0}$.

\section{The interpolation parameter}
\label{sec:theta}

The analysis of the previous section has some interest as an alternative procedure to obtain the already known relationships derived by Darken (for the limits and range of application of these equations, see, e.g., \cite{mehrer}). Nevertheless, the main purpose of this paper is to further advance in the description of diffusion in binary alloys using the mean field approach. In order to do that, in this section we obtain an expression of the interpolation parameter $\theta_A$ as a function of the mean field potential $V_A$ and, using this result, in the next section we demonstrate that the intrinsic diffusivity does not depend on the thermodynamic factor. In the rest of this section we omit subindices $A$ and $i$ to lighten the notation, and assume that we deal with $n$ particles of species $A$ in cell number $i$; the cell has volume $v$.

The first step is to find the connection between the mean field potential $V$ and the energy $\phi$ of $n$ particles in a cell. We use the local equilibrium assumption to consider that the energy is a function only of the number of particles and the temperature. We can write the grand partition function of the cell as
\begin{equation}
\mathcal{Z} = \sum_{n=0}^{\infty} \frac{1}{n!} e^{-\beta[\phi(n) - \tilde{\mu} n]},
\end{equation}
where $\tilde{\mu} = \mu/\mathcal{N}_A$ is the chemical potential per particle and $\mathcal{N}_A$ is the Avogadro's constant.
The mean number of particles is
\begin{align}
\bar{n} &= \frac{1}{\beta} \dpar{\ln \mathcal{Z}}{\tilde{\mu}} = \frac{1}{\mathcal{Z}} \sum_{n=0}^{\infty} \frac{n}{n!} e^{-\beta[\phi(n) - \tilde{\mu} n]} \nonumber \\
&= \frac{e^{\beta \tilde{\mu}}}{\mathcal{Z}} \sum_{n=1}^{\infty} \frac{1}{(n-1)!} e^{-\beta[\phi(n) - \tilde{\mu} (n-1)]} \nonumber \\
&=  \frac{e^{\beta \tilde{\mu}}}{\mathcal{Z}} \sum_{m=0}^{\infty} \frac{1}{m!} e^{-\beta[\phi(m+1) - \tilde{\mu} m]} \nonumber \\
&=  \frac{e^{\beta \tilde{\mu}}}{\mathcal{Z}} \sum_{m=0}^{\infty} \frac{1}{m!} e^{-\beta[\phi(m+1)-\phi(m)]} e^{-\beta[\phi(m) - \tilde{\mu} m]} \nonumber \\
&= e^{\beta \tilde{\mu}} \langle e^{-\beta[\phi(n+1)-\phi(n)]} \rangle,
\label{eq:nmed1}
\end{align}
where, in the third line, we changed the summation index: $m=n-1$. The concentration is $c=\bar{n}/v$. Using Eqs.\ \eqref{eq:conc} and \eqref{va1} (with subindex $A$ omitted), we have
\begin{equation}
\bar{n} = e^{\beta \tilde{\mu}} e^{-\beta(V+b)}
\label{eq:nmed2}
\end{equation}
where $b$ is a constant given by $b = \tilde{\mu}^0 - k_B T \ln (c_{B0}v/\gamma_0)$, and $k_B=R/\mathcal{N}_A$ is the Boltzmann constant. Comparing Eqs.\ \eqref{eq:nmed1} and \eqref{eq:nmed2}, we get
\begin{equation}
e^{-\beta(V+b)} = \langle e^{-\beta\,\Delta \phi(n)}\rangle
\label{eq:Vbphi}
\end{equation}
with $\Delta \phi(n) = \phi(n+1)-\phi(n)$, and where $V$ is evaluated in $\bar{n}$. As usual in thermodynamics, we treat $\phi$ as a continuous function of $n$. We use the following notation: $\phi$ without explicit dependence means that it is evaluated in $\bar{n}$, and we use prime to represent derivatives respect to $\bar{n}$: $\phi'= \frac{d\phi}{d\bar{n}}$. We can approximate the average of the previous equation and obtain an expression in terms of the mean square fluctuations of the number of particles that, in turn, can be evaluated with the partition function. The result is (see Appendix B for the details of this derivation)
\begin{equation}
\phi' = V + b -\frac{1}{2\beta} \frac{d\ln(1+\beta \bar{n} V')}{d\bar{n}}.
\label{eq:phiVb}
\end{equation}

The next step is to find an equation that connects the interpolation parameter $\theta$ with $V$ and $\phi$. Let us consider a process in which one particle jumps from cell number 1 to cell number 2. The rest of the cells remain unchanged, so we specify the configuration of the system using only the number of particles in cells 1 and 2. Initially we have the configuration given by $\{n_1,n_2\}$; after the jump we have $\{n_1-1,n_2+1\}$. In order to adopt a more compact notation, we now use $V_{n_i}$ instead of $V(n_i)$, and the same for $\phi(n_i)$ and $\theta(n_i)$.  From the exponential in the transition probability \eqref{eq:transition}, we have the height of the energy barrier that has to be overcome to perform the jump:
\begin{equation}
h_\text{ini} = \frac{\theta_{n_1}-1}{2} V_{n_1} + \frac{\theta_{n_2}+1}{2} V_{n_2}.
\end{equation}
If we consider the reverse process $\{n_1-1,n_2+1\} \rightarrow \{n_1,n_2\}$, the energy barrier is
\begin{equation}
h_\text{fin} = \frac{\theta_{n_2+1}-1}{2} V_{n_2+1} + \frac{\theta_{n_1-1}+1}{2} V_{n_1-1}.
\end{equation}
\begin{figure}
	\includegraphics{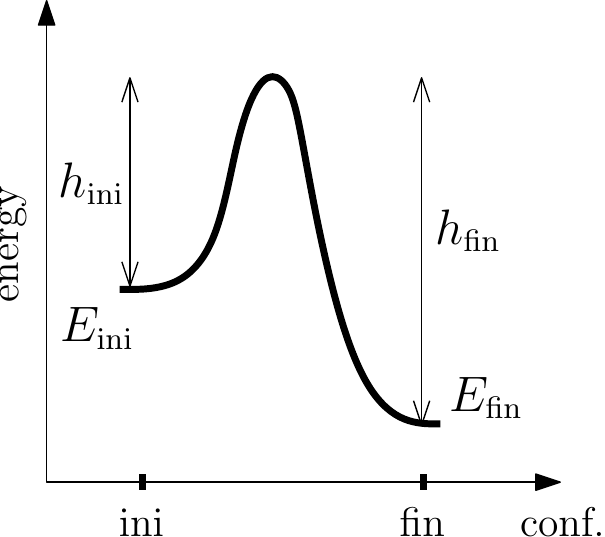}
	\caption{Diagram of the energy against configuration when a particle jumps between cells 1 and 2. The initial configuration corresponds to $\{n_1,n_2\}$ and the final configuration to  $\{n_1-1,n_2+1\}$.}
	\label{confdiag}
\end{figure}
From Figure \ref{confdiag} we can see that the energy difference between configurations is
\begin{equation}
\Delta E = E_\text{fin} - E_\text{ini} = h_\text{ini} - h_\text{fin}.
\end{equation}
On the other hand, using the function $\phi$ for the energy of a given number of particles in a cell, we have that the energy difference is
\begin{equation}
\Delta E = \phi_{n_1-1} + \phi_{n_2+1} - \phi_{n_1}- \phi_{n_2}.
\end{equation}
Combining the last equations, and rearranging terms, we have
\begin{align}
&\phi_{n_2+1} - \phi_{n_2} - \frac{V_{n_2+1}+V_{n_2}}{2} + \frac{\theta_{n_2+1} V_{n_2+1} - \theta_{n_2} V_{n_2}}{2} \nonumber \\
&= \phi_{n_1} - \phi_{n_1-1} - \frac{V_{n_1}+V_{n_1-1}}{2} + \frac{\theta_{n_1} V_{n_1} - \theta_{n_1-1} V_{n_1-1}}{2}.
\end{align}
Now we perform a series expansion of $\phi$ and $V$ around $\bar{n}_2$ in the left hand side and $\bar{n}_1$ in the right hand side; it is the same kind of expansion that is used in Appendix B, where we keep terms up to order $v^{-1}$ and the volume $v$ is used as a large parameter.  After taking the average on different realizations, we obtain
\begin{equation}
\left(\phi' - V +\frac{1}{2} \frac{d \theta V}{d \bar{n}} \right)_{\bar{n}_2} = \left(\phi' - V +\frac{1}{2} \frac{d \theta V}{d \bar{n}} \right)_{\bar{n}_1}.
\end{equation}
There is a small and arbitrary concentration difference between sites 1 and 2, therefore, the previous equation implies
\begin{equation}
\frac{d}{d \bar{n}}\left(\phi' - V +\frac{1}{2} \frac{d \theta V}{d \bar{n}} \right) = 0.
\end{equation}
We use Eq.\ \eqref{eq:phiVb} for $\phi'$, cancel constant $b$ and obtain
\begin{equation}
\frac{d^2}{d \bar{n}^2} \left[ \beta \theta V - \ln (1+\beta \bar{n} V') \right]=0,
\end{equation}
or
\begin{equation}
\beta \theta V - \ln (1+\beta \bar{n} V') = \kappa_1 \bar{n} + \kappa_2.
\end{equation}
Constants $\kappa_1$ and $\kappa_2$ are obtained from the following conditions. In the limit of small concentration, $\bar{n}\rightarrow 0$, we have that $V=0$, see  Eq.\ \eqref{va1}. This condition implies that $\kappa_2=0$. Now, using that $\lim_{\bar{n}\rightarrow 0}V/\bar{n}= V'$, we can write
\begin{align}
\kappa_1 &= \beta \theta V' - \frac{1}{\bar{n}}\ln (1+\beta \bar{n} V') \nonumber \\
& \simeq \beta \theta V' - \beta V' \qquad \qquad \qquad (\bar{n} \rightarrow 0)\nonumber \\
&= 0
\end{align}
where we have used the condition that $\theta=1$ for $\bar{n}\rightarrow 0$. It is deduced from the interpretation of $\theta$ as an interpolation parameter that determines if the transition probability $W_{i,i+1}$ [see Eq.\ \eqref{eq:transition}] depends on the potential in the origin or target cell. For small concentration there is at most one particle in a cell, and the probability to jump to a neighboring cell depends on whether it is occupied or not by another particle, that is, it depends on the potential in the target cell. This means that in the limit of small concentration we have $\theta=1$.

The final result is
\begin{equation}
\beta \theta V = \ln(1+\beta \bar{n} V').
\label{eq:thetaV}
\end{equation}

\section{Dependence on the thermodynamic factor}
\label{sec:thermofactor}

In this section, we recover the more specific notation with subindex $A$ to specify the type of component. The result obtained for $\theta_A$, Eq.\ \eqref{eq:thetaV}, is directly related to the thermodynamic factor. It can be shown that
\begin{equation}
1+\beta \bar{n}_A V'_A = 1 + \beta c_A \dpar{V_A}{c_A}
= \Phi \frac{\nu_m}{\nu_{B}},
\label{eq:tfact}
\end{equation}
where $c_A = \bar{n}_A/v$, and we have used part of the calculations presented in Appendix A. Then, using \eqref{eq:tfact} in \eqref{eq:thetaV}, we get
\begin{equation}
e^{-\beta \theta_A V_A} = \Phi^{-1} \frac{\nu_{B}}{\nu_m},
\end{equation}
and using this last result in the equations for the intrinsic and tracer diffusivities, Eqs.\ \eqref{eq:intr} and \eqref{eq:trac}, we finally obtain:
\begin{align}
D_A &= D_{A 0}\, e^{-\beta (g_A-g_{A 0})} \label{DA}  \\
D_A^* &= D_{A 0}\, e^{-\beta (g_A-g_{A 0})}  \,\Phi^{-1} \frac{\nu_{B}}{\nu_m} , \label{DAa}
\end{align}
where we have used \eqref{eq:Delta} for $\Delta_A$.

The thermodynamic factor $\Phi$ depends on the activity coefficient, that represents the departure from the behavior of an ideal mixture due to interactions between $A$ and $B$ species. The previous analysis shows that the intrinsic diffusivity does not depend on the thermodynamic factor, and it is mainly determined by the activation energy $g_A$. On the other hand, the tracer diffusivity behaves as $\Phi^{-1}$.

A first approximation for the mole fraction dependence of the activation energy is
\begin{equation}
g_A = N_A g_{A 1} + N_{B} g_{A 0} - \varepsilon_{A} N_A N_{B},
\label{eq:gA}
\end{equation}
where $g_{A 1}$ and $g_{A 0}$ are the activation energies for $N_A \rightarrow 1$ and $N_A \rightarrow 0$, respectively. The first two terms in the previous equation represent the Vegard's law. The last term is a possible departure from Vegard's law; it has the same shape as the correction term in the mixing energy that gives rise to Margules equations (see, e.g., \cite[p. 150]{atkins}). For more elaborate representations of the vacancy formation energy, included in $g_A$, see, e.g., \cite{bakker,kim}; for an introduction to defect-mediated diffusion, see \cite[ch.\ 10]{paul2}. Now, the intrinsic diffusivity takes the form
\begin{equation}
D_A = D_{A 0} \, e^{-\beta \,\Delta g_A\,N_A}\, e^{\beta \varepsilon_{A} N_A N_{B}}
\end{equation}
with $\Delta g_A = g_{A 1}-g_{A 0}$. Using the value of the intrinsic diffusivity in the limit of $N_A\rightarrow 1$, $D_{A1} = D_{A 0} \, e^{-\beta \,\Delta g_A}$, we have
\begin{equation}
D_A = D_{A 0}^{N_B} D_{A1}^{N_A}\, e^{\beta \varepsilon_{A} N_A N_{B}}.
\label{eq:difequ}
\end{equation}

\section{Experimental test}
\label{sec:exptest}

\begin{figure}
\includegraphics[width=0.48\textwidth]{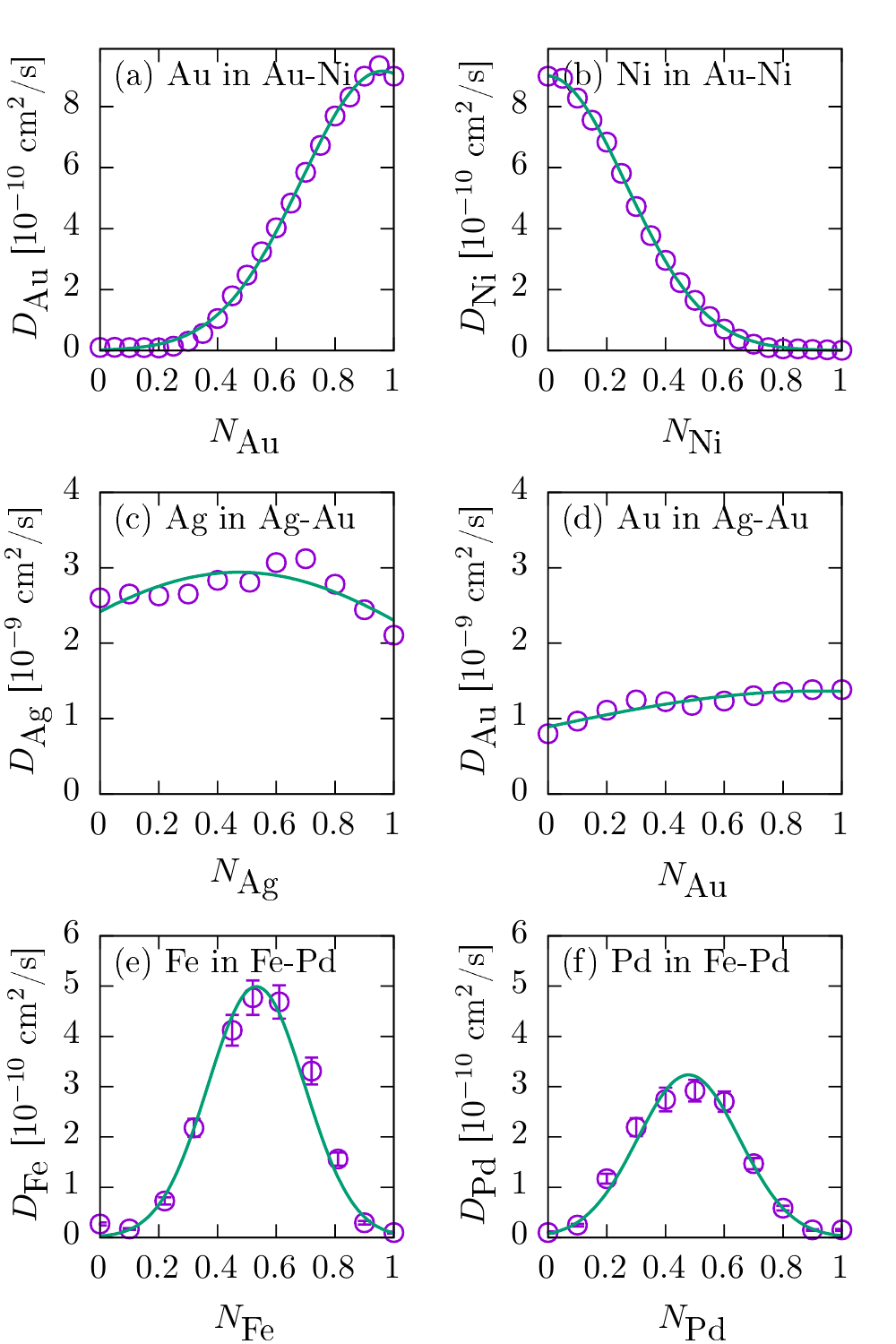}
	\caption{Intrinsic diffusivities as a function of the molar fraction: Experimental data extracted from Refs.~\cite{reynolds,mead,fillon} (circles) and the Eq.~\eqref{eq:difequ} fitted for each set (line). See Table \ref{tab:fit} for the parameters used in each case.}
	\label{fig:dif}
\end{figure}

To test the expression found for the intrinsic diffusivity in the mean-field theory approach, we gather previous experimental data and fit them with Eq.~\eqref{eq:difequ} for different metal alloys. In Fig.~\ref{fig:dif}, we show these results for Au-Ni \cite{reynolds}, Ag-Au \cite{mead} and Fe-Pd \cite{fillon} and in Table~\ref{tab:fit} we present the fitted parameters $D_{0}$, $D_{1}$ (diffusivities for molar fractions in the limits 0 and 1; subindex $A$ is omitted for simplicity), with their respective reference values for comparison, and $\beta\varepsilon$.

We found that the qualitatively different behaviors shown in Fig.\ \ref{fig:dif} can be reproduced, with good agreement, using the expression of Eq.\ \eqref{eq:difequ}, with an appropriate fit of parameters $D_0$, $D_1$ and $\beta\epsilon$.

\begin{table}[htbp]
	\centering
	\begin{tabular}{@{}p{0.08\textwidth}*{5}{L{\dimexpr0.08\textwidth-2\tabcolsep\relax}}@{}}
		\toprule
		& \multicolumn{2}{c}{$D_{0}$} &
		\multicolumn{2}{c}{$D_{1}$} & \multicolumn{1}{c}{$\beta\varepsilon$}\\
		\cmidrule(r{4pt}){2-3} \cmidrule(l){4-5}
		& Our result & Ref. value & Our result & Ref. value & \\
		\midrule
		Au & $0.022(4)$ & $0.1$ & $9.1(1)$ & $9$ & $6.5(3)$\\
		 &  &  $0.015$ &  & $14.8(6)$ & \\
		Ni & $9.01(5)$ & $9$ & $0.009(1)$ & $0.006$ & $6.9(2)$\\
		&  & $8$ & & -- & \\
		\midrule
		Ag & $24(1)$ & $26$ & $23(1)$ & $21$ & $0.9(3)$\\
		&  & $21$ &  & $22(3)$ & \\
		Au & $8.9(5)$ & $8$ & $13.6(5)$ & $13.8$ & $0.5(2)$\\
		&  & $7.6$ &  & $13.5(5)$ & \\
		\midrule
		Fe & $0.03(1)$ & $0.27(3)$ & $0.09(2)$ & $0.10(2)$ & $18(1)$\\
		&  & $0.23$ &  & $0.06$ & \\
		Pd & $0.07(2)$ & $0.10(2)$ & $0.03(1)$ & $0.15(2)$ & $16(1)$\\
		&  & $0.08$ &  & $0.15$ & \\
		\bottomrule
	\end{tabular}
	\caption{Parameters obtained by fitting Eq.~\eqref{eq:difequ} for each metal diffusing in their respective alloy. Reference values for $D_{0}$ and $D_{1}$ are shown for comparison. These were extracted from Ref.~\cite{reynolds} for the Au-Ni alloy, Ref.~\cite{mead} for Au-Ag and Ref.~\cite{fillon} for Fe-Pd. Extra reference values were calculated using experimental data from Refs.~\cite{sohn,neumann,peterson}. Units for $D_0$ and $D_1$ are $10^{-10}$cm$^2$/s.}
	\label{tab:fit}
\end{table}

Table \ref{tab:fit} also shows reference values of $D_0$ and $D_1$ taken from Refs.~\cite{reynolds,mead,fillon}. The agreement with the fitting parameters that we obtained is good except in cases in which the diffusivity takes small values and the relative error is larger.
In addition, extra reference values are presented for $D_0$ and $D_1$. These were calculated using experimental data from Refs.~\cite{sohn,neumann,peterson}, where different contributions to self-diffusion, like monovacancy, divacancy and vacancy migration, are taken into account.
Errors are reported when available.

There is a physical argument to qualitatively understand why the values obtained for $\epsilon$ are positive. As mentioned in Sect.\ \ref{sec:mean}, the activation energy is the sum of the migration energy plus the vacancy formation energy. The mixture of two species with different properties in a solid alloy creates disorder in an otherwise ordered lattice (for a pure material). The disorder favors the formation of vacancies, therefore the vacancy formation energy should be smaller than the linear interpolation represented by the Vegard's law (as long as the molar fraction takes values different from 0 or 1). This is reflected by the negative nonlinear term in the activation energy, Eq.\ \eqref{eq:gA}, i.e., a positive value of $\epsilon$.

\section{Conclusions}
\label{sec:conclusions}

Starting from the expression of the transition probabilities \eqref{eq:transition} in terms of the mean field potential $V_A$ and the interpolation parameter $\theta_A$ (introduced in Ref.\ \cite{suarez}) we can obtain an alternative derivation of the Darken equation \eqref{dark1}. More interesting, we obtain separate expressions for the intrinsic and tracer diffusion coefficients, Eqs.\ \eqref{DA} and \eqref{DAa}. The factors that determine the dependence of the diffusivities on concentration are the thermodynamic factor $\Phi$ and the activation energy $g_A$. We obtained that the intrinsic diffusivity $D_A$ does not depend on $\Phi$. We use a quadratic form for the dependence of $g_A$ on the molar fraction, Eq.\ \eqref{eq:gA}. The resulting expression for the intrinsic diffusivity, Eq.\ \eqref{eq:difequ}, has three parameters: the diffusivity in the limits of molar fraction 0 and 1, and the coefficient $\epsilon_A$ of the quadratic term of the activation energy. By fitting these parameters, we show that Eq.\ \eqref{eq:difequ} is able to correctly represent experimental results of the intrinsic diffusivity for three different metal alloys, see Fig.\ \ref{fig:dif}. The available experimental data found in the literature are consistent with the theory here developed, and this is a promising result. However, in order to have a thorough test of the theory it is still necessary, for example, to have access to accurate experimental values of the activation energy, or to consider systems with a concentration dependence of the thermodynamic factor whose influence can be observed in $D_A^*$ and not in $D_A$. Besides the activation energy and the thermodynamic factor, there are other elements at stake to be considered in a more detailed description, and it is experimentally challenging to discriminate all of them. For example, the vacancy wind factor or Manning factor, the presence of impurities, or the impurity vacancy binding energy; see \cite{santra}.

%Factors to be considered for further refinements: a more accurate description of the number of vacancies against species mole fraction; the probability to find a vacancy may differ for different kinds of particles; inclusion of the vacancy wind factor or Manning factor; presence of impurities, and impurity vacancy binding energy; see \cite{santra}.

\section*{Appendix A}
%\begin{align}
%\beta c_A \dpar{V_A}{x} + \dpar{c_A}{x} &= c_A \frac{\partial}{\partial x} \left( \beta  V_A + \ln c_A\right) \nonumber \\
%&= c_A \frac{\partial}{\partial x} \ln(\gamma_A c_A/c_T) \nonumber \\
%&= c_A \frac{\partial}{\partial x} \ln(\gamma_A N_A)
%\end{align}
%where we have used Eq.\ \eqref{va1} for the mean field potential $V_A$.

We present here more details of the derivation of Eq.\ \eqref{eq:Jalpha} from Eq.\ \eqref{corrJ}. Let us focus attention on the parenthesis in the right hand side of \eqref{corrJ}, and let us call it $X$ for further reference:
\begin{align}
X &= \beta c_A \dpar{V_A}{x} + \dpar{c_A}{x} \nonumber \\
&= \left(\beta c_A \frac{\partial V_A}{\partial c_A} + 1 \right) \dpar{c_A}{x} \nonumber \\
&= \left( c_A \frac{\partial \ln (\gamma_A/c_T)}{\partial c_A} +1 \right) \dpar{c_A}{x} \nonumber \\
&= \left( c_A \frac{\partial \ln \gamma_A}{\partial N_A} \frac{d N_A}{d c_A} - \frac{c_A}{c_T} \frac{d c_T}{d c_A} +1 \right) \dpar{c_A}{x}
\label{eq:X}
\end{align}
where we have used Eq.\ \eqref{va1} for the mean field potential $V_A$.
For partial molar quantities, as $\nu_A$ and $\nu_B$, the following relation holds: $\nu_A \, dc_A + \nu_B \,dc_B=0$. Then, since $c_T = c_A + c_B$, we have
\begin{equation}
\frac{dc_T}{dc_A} = 1-\frac{\nu_A}{\nu_B}
\label{dctdca}
\end{equation}
and, using that $N_A=c_A/c_T=c_A \nu_m$,
\begin{align}
\frac{d N_A}{d c_A} &= \nu_m - \frac{c_A}{c_T^2} \frac{d c_T}{d c_A} \nonumber \\
&= \left( \nu_m - c_A \nu_m^2(1-\nu_A/\nu_B) \right) \nonumber \\
&= \nu_m^2(c_T - c_A + c_A \nu_A/\nu_B) \nonumber \\
&=  \frac{\nu_m^2}{\nu_B} (c_B \nu_B + c_A\nu_A) \nonumber \\
&=  \frac{\nu_m^2}{\nu_B}.
\label{dNa}
\end{align}
Using \eqref{dctdca} and \eqref{dNa} in \eqref{eq:X}, we have
\begin{align}
X&= \left[ c_A \frac{\partial \ln \gamma_A}{\partial N_A} \frac{\nu_m^2}{\nu_B}  - N_A \left( 1-\frac{\nu_A}{\nu_B}\right) +1 \right] \dpar{c_A}{x} \nonumber \\
&= \left( \frac{\nu_m}{\nu_B} \frac{\partial \ln \gamma_A}{\partial\ln N_A} + N_B + N_A \frac{\nu_A}{\nu_B}  \right) \dpar{c_A}{x} \nonumber \\
&= \left( \frac{\partial \ln \gamma_A}{\partial\ln N_A} + 1 \right) \frac{\nu_m}{\nu_B} \dpar{c_A}{x} \nonumber \\
&= \Phi \frac{\nu_m}{\nu_B} \dpar{c_A}{x},
\end{align}
where, in the last step, we have used the definition of the thermodynamic factor $\Phi$ \eqref{eq:Phi}. Replacing this expression for $X$ in \eqref{corrJ}, we obtain \eqref{eq:Jalpha}.

\section*{Appendix B}

In this appendix we derive Eq.~\eqref{eq:phiVb} from Eq.~\eqref{eq:Vbphi}.
The relevant values of $\Delta \phi(n)$, in the average of Eq.~\eqref{eq:Vbphi}, are similar to $\phi'$, so we can approximate
\begin{align}
e^{-\beta(V+b)} &= e^{-\beta \phi'} \langle e^{-\beta[\Delta \phi(n) - \phi']}\rangle \nonumber \\
&= e^{-\beta \phi'}[ 1 - \beta\langle \Delta \phi(n) - \phi'\rangle \nonumber \\
&\ \ \ + \beta^2 \langle (\Delta \phi(n) - \phi')^2\rangle/2 + \cdots ].
\label{eq:apB}
\end{align}
For, for example, $\phi(n)$ we can write
\begin{equation}
\phi(n)=\phi(\bar{n}+\Delta n) = \phi + \phi'\,\Delta n + \frac{1}{2} \phi''\, \Delta n^2 + \cdots
\end{equation}
where $\Delta n = n - \bar{n}$. Using an expansion of $\Delta \phi(n)$ in terms of the fluctuation $\Delta n$ in Eq.\ \eqref{eq:apB}, we obtain
\begin{equation}
e^{-\beta(V+b)} = e^{-\beta \phi'}(1 + \beta \epsilon)
\label{eq:Vbeps}
\end{equation}
with
\begin{equation}
\epsilon = -\frac{1}{2}\phi'' + \frac{1}{2}(\beta\phi''^2 - \phi''') \langle\Delta n^2\rangle.
\label{eq:epsphi}
\end{equation}
The mean number of particles $\bar{n}$ is an extensive quantity, proportional to the volume $v$; we have that $\phi'' \sim v^{-1}$, $\phi''' \sim v^{-2}$ and so on, and  $\langle\Delta n^2\rangle \sim v$ (we can check this bellow). Then, $\epsilon$ is of order $v^{-1}$ and we have neglected terms of order $v^{-2}$ or smaller in \eqref{eq:epsphi}. Applying logarithm to Eq.\ \eqref{eq:Vbeps} we get
\begin{equation}
\phi' = V + b + \epsilon,
\end{equation}
and, deriving with respect to $\bar{n}$,
\begin{align}
\phi'' &= V' + \mathcal{O}(v^{-2}) \\
\phi''' &= V'' + \mathcal{O}(v^{-3}).
\end{align}
So, we can rewrite $\epsilon$, keeping the same degree of accuracy, as
\begin{equation}
\epsilon = -\frac{1}{2}V' + \frac{1}{2}(\beta V'^2 - V'') \langle\Delta n^2\rangle.
\label{eq:epsV}
\end{equation}
We obtain the average of the squared fluctuations from
\begin{equation}
\langle\Delta n^2\rangle = \frac{1}{\beta^2} \frac{\partial^2 \ln \mathcal{Z}}{\partial \tilde{\mu}^2} = \frac{1}{\beta} \dpar{\bar{n}}{\tilde{\mu}}.
\end{equation}
Let us notice that from this equation we can obtain the known relationship for the thermodynamic factor against fluctuations \cite[Sec.\ 2.6]{gomer}:
\begin{equation}
\frac{\nu_m}{\nu_B} \Phi = \beta \dpar{\tilde{\mu}}{\ln c_A} = \beta \bar{n} \dpar{\tilde{\mu}}{\bar{n}} = \frac{\bar{n}}{\langle \Delta n^2\rangle}.
\end{equation}
Using Eq.\ \eqref{eq:nmed2} to obtain $\dpar{\bar{n}}{\tilde{\mu}}$, we get
\begin{equation}
\langle\Delta n^2\rangle = \frac{\bar{n}}{1 + \beta \bar{n} V'}
\end{equation}
and, replacing in \eqref{eq:epsV}, we have
\begin{equation}
\epsilon = -\frac{V' + \bar{n}V''}{2(1 + \beta \bar{n} V')}=-\frac{1}{2\beta} \frac{d\ln(1+\beta \bar{n} V')}{d\bar{n}}.
\end{equation}
Finally, this result for $\epsilon$ gives us Eq.\ \eqref{eq:phiVb}.

\begin{acknowledgments}
The authors acknowledge H. M\'artin for illuminating discussions. This work was partially supported by Consejo Nacional de Investigaciones Cient\'ificas y T\'ecnicas (CONICET, Argentina, PIP 112 201501 00021 CO).
\end{acknowledgments}

\bibliography{difequ.bib}

\end{document}